\documentclass[preprint,double-spaced]{JHEP3}

\usepackage{graphicx}
\usepackage{dcolumn}
\usepackage{bm}

\title{Delineating cosmic expansion history with recent supernova data: \\
A Bayesian model-independent approach}

\author{Moncy V. John, \\
Department of Physics, St. Thomas College, 
Kozhencherry, Kerala - 689 641, India \\
\email{moncy@iucaa.ernet.in, moncyjohn@yahoo.co.uk} }

\date{\today}

\abstract{
 Marginal likelihoods for the cosmic expansion rates are evaluated using the recent `Constitution'  data of 397 supernovas, thereby updating the results  in some previous works. Even when beginning with a very strong prior probability that favors an accelerated expansion, we end up with a marginal likelihood for the deceleration parameter $q_0$ peaked around zero in the spatially flat case. This is in agreement with some other analysis of the Constitution data. It is also  found that the new data  significantly constrains the cosmic expansion rates, when compared to the previous analyses. Here again we adopt the model-independent approach  in which  the scale factor is expanded into a Taylor series in time about the present epoch; for practical purposes,  it is truncated to    polynomials of various orders, in different trials. Though one cannot regard the polynomials thus obtained as   models, in this paper we  evaluate the total likelihoods (Bayesian evidences) for them  to find the order of the polynomial having the largest  likelihood.  Analysis using the  Constitution data shows that the largest likelihood  occurs for the fourth order polynomial and is of value $\approx 0.77 \times 10^{-102}$. It is argued that this value, which we call the likelihood for the model-independent approach, may be used to calibrate the performance of realistic models. 
}


\begin{document}

\section{Introduction}
A decade ago, the apparent magnitude-redshift ($m-z$) data of  type Ia supernova (SN Ia)    were declared to be indicating that our universe is expanding with an acceleration \cite{per99,rie98}. This discovery is considered to be the most important and startling one in cosmology, after the 1929 observation by Hubble that the universe is expanding, and the 1964 discovery by Penzias and Wilson that there is a  cosmic microwave background. The discovery of accelerated expansion is startling for it demands the presence of substantial amount of some unknown `dark energy' in the universe, with repulsive pressure. For the analysis of the data, both the  supernova search teams which first reported this observation assumed the validity of the LCDM model in cosmology, in which the universe  contains  a cosmological constant $\Lambda$ or an equivalent vacuum energy $\rho_{\Lambda}$ (with an equation of state $p_{\Lambda}=-\rho_{\Lambda}$), along with  nonrelativistic matter. The latter   includes some  unknown cold `dark matter' too. Analyses were also performed under the  speculation that the  dark energy has the equation of state $p_{de}=w\rho_{de}$, with $w$ as a time-varying quantity. The Chevallier-Polarski-Linder (CPL) ansatz \cite{cpl1,cpl2} is the popular one in this connection. Later it was  speculated that there is  interaction between the energy components \cite{pavon,intrde}. It may even be asked whether there is only one component in the `dark energy' at the present epoch \cite{gong}.

We may  note that in all the above  analyses, the following assumptions are made: (1) The Einstein equation in general relativity   is valid for the cosmos (which are the Friedmann equations). (2) The universe  is homogeneous and isotropic at very large scales and hence spacetime possesses a Robertson-Walker (RW) metric. (3) There can be  one or more components in the  cosmic fluid and in the latter case, the components may or may not be separately conserved. (However, the total energy density will be  conserved.) Specific assumptions in this regard, on each component and their equations of state, are  prerequisites to begin the analysis, for we need a solution $a(t)$ of   the Einstein equation in this case (where $a(t)$ is the scale factor of expansion) at our disposal. Such assumptions and the resulting solution $a(t)$ constitute  a `cosmological model'. It shall be noted that $a(t)$ often contains some free parameters, but  the specific values assumed by these parameters alone do not lead to different models, unless those values  are predictions in them, on the basis of some fundamental principles. 

When we make such model-based analyses, one of the tasks is to find the model which enjoys the largest support from the data.  The merit of a new model is  adjudged by comparing it with any existing ones, using the data related to the phenomena under consideration. For this,  the conventional approach is to perform a maximum likelihood ratio test  where one has to find the lowest $\chi^2$  for the models concerned. But  it is now recognized that the use of  Bayes theory is the more reasonable approach in problems such as  cosmology \cite{jaffe,drell,mvjvn,hobson}.  In Bayesian model comparison, we take  two  models as rival hypotheses  (say, $M_i$ and $M_j$) and try to compare them by evaluating odds ratios between their posterior (i.e., after analyzing the data) probabilities, given the data $D$ and  also some background information $I$. To obtain these ratios, the Bayes's theorem is made use of. In the cosmological  literature,  a large amount of work which use Bayes theorem is now reported [See \cite{pia,trotta} for some  reviews]. But  a clear distinction between Bayesian parameter estimation and Bayesian model comparison is not found in  most of the reported works. It shall here be noted that cosmological models which are  truly distinct (in the  sense of the definition of the term given above) were first compared using  Bayesian theory  in \cite{mvjvn}.

 We  note that since there can be infinitely many solutions  $a(t)$ based on our specific assumptions regarding energy densities, it is not possible to  evaluate  any one model's  posterior probability. Instead, as mentioned above, one  can  perform a model comparison. Comparison is possible because here we need only to find the ratios between   the probabilities for obtaining the data $D$ in the various models,  multiplied by any prior odds. The probability for obtaining the data $D$ in a model is often referred to as the `likelihood for the model'  or the `Bayesian evidence'. 

On another front, it was pointed out that the most appropriate way to measure the acceleration of the universe is to resort to a model-independent approach. In the conventional model-based analyses of $m-z$ data of SN Ia,  the accelerated expansion of the universe was  an indirect inference based on the best fit values of parameters, such as the density parameters $\Omega_m$ and $\Omega_{\Lambda}$ in the LCDM model.  In the model-independent case, the scale factor $a(t)$ is  expanded as a Taylor series in time about the present epoch \cite{mvj1,mvj2} and the marginal likelihoods of its coefficients are computed using the data. But practically, we have to truncate the series to some finite order and hence our basic assumption is that $a(t)$ is expressible as a truncated Taylor series or polynomial. Evaluating the deceleration parameter  by adopting this method, it was  confirmed model-independently using the SN data that the  universe is undergoing an accelerated expansion \cite{mvj1}. Clearly, this Taylor expanded scale factor is not  a realistic cosmological  model, since here  homogeneity and isotropy  are the only assumptions made  for the universe.

Model-independent approaches which use Taylor expansion in terms of redshift $z$ have also gained wide attention in recent years [See for eg. \cite{turner,visser,lima,seikel}]. A great advantage of the present approach of expanding the scale factor  in  terms of $t$ about the present epoch $t_0$, when compared to expansion in   $z$ about $z=0$ is that the former converges for all times, whereas the latter converges only for $\mid z\mid <1$ \cite{visser}. In the present paper, even the lookback time $T\equiv t-t_0$ is evaluated by numerically solving Eq. (13) in Ref. \cite{mvj2}, which involves a  Taylor series in time; here we do not use an expansion in $z$ at all. Hence there is no convergence problem \cite{visser,lima} in the present work.

It was mentioned above that in  Bayesian model comparison one  evaluates only the relative merits of models on the basis of  data and  it is not  possible to compute the posterior probability for a model. Also,  Bayesian model comparison  does not give  the best possible model, for it only compares the  available ones.
In this circumstance we ask whether one can find some standard, which can be considered as a minimum requirement, for a model to be termed  successful. In the lowest $\chi^2$ per degree of freedom approach, such a crude standard exists; a model is considered a reasonable good fit if the $\chi^2$ per degree of freedom is less than or nearly equal to unity. Several analyses of SN data in cosmology still use this standard [for eg. \cite{varun,wei}]. We attempt in this paper  to set a similar standard in Bayesian analysis. Obviously, the relevant  quantity in this context is   the  likelihood for the model or the Bayesian evidence. We  propose to evaluate this quantity for the scale factor $a(t)$ in its Taylor series form about the present epoch, thus combining the Bayesian and model-independent approaches. Eventhough one cannot regard the truncated Taylor series  form of the scale factor as a model,   the likelihood for polynomials of various orders may be obtained, and thereby one can obtain the order of the polynomial  with the largest likelihood. That such a maximum exists can be seen from the fact that the Bayesian approach has a built-in mechanism  to implement the Occam's razor; i.e., it favors simpler models when compared to complicated ones with more parameters, unless the latter shows significantly better performance. We have assumed that the scale factor of the universe has a unique Taylor expansion with definite values for its coefficients, which we attempt to find with the help of data. Varying the coefficients in the  series arbitrarily will certainly affect $\chi^2$; they are not unconstrained parameters. When the order of the truncated Taylor series becomes very large, the number of coefficients in it too becomes  large so that Occam's razor forces the likelihood of the polynomial to tend to zero. Therefore it should be possible to find the  order of the polynomial that maximizes the likelihood. We here argue that this maximum value of the likelihood/evidence for the Taylor expansion, which may be termed  the likelihood for the model-independent approach, can be set as a standard for model comparison. 

It shall here  be noted that a great practical use of scientific theories is that the equations they provide save us from  keeping large amount of raw data, for use in future  applications. Therefore it is natural to expect that the likelihood for   a fundamental scientific theory  exceed that of a truncated Taylor series of the unknown function  or at least be equal to that of such a series.  In other words, only those realistic models, which have likelihoods greater than or at least equal to that in the model-independent approach can be considered as  successful. Even in cases where it is not possible to find the order of the polynomial that maximises the likelihood for some practical reason, it is reasonable to demand that realistic models   perform better than each of the (low order) polynomials we have worked out.

Bayesian model-independent approaches of the  kind proposed here are  pursued by some authors, though not in a  systematic way as in \cite{mvj2} or as proposed to do in this paper. For instance, Guimaraes, Cunha and Lima \cite{lima} have compared the realistic  flat LCDM model with simple kinematic models, such as those based on three simple parametrisations for the deceleration parameter etc., using the 307  SN Ia Union compilation  data set. They found that even very simple kinematic models are equally good to describe this data, when compared to LCDM model. We develop this procedure further and attempt to achieve  a systematic calibration of cosmological models, by first evaluating the likelihoods for polynomials of various orders and then finding the largest possible value of the  likelihood for a Taylor expansion. For calibration purpose, we compare other model likelihoods with this value using the Bayesian method. Moreover, in this work we use a more recent  `Constitution' supernova data \cite{hicken} which contains 397 objects. 

  Another important work we report in this paper is that of   updating the  marginal likelihood for each of the expansion coefficients found in \cite{mvj1,mvj2}. This is performed for the case of a fifth order polynomial. The new marginal likelihoods  for its coefficients  give valuable information regarding the expansion history of the universe. An interesting result in this connection is that even when beginning with a very strong prior probability that favors an accelerated expansion, we end up with a marginal likelihood for the deceleration parameter $q_0$ peaked around zero in the spatially flat case. This result is in agreement with some other analyses [See for eg., \cite{varun}]. It is also  found that the new data  significantly constrains the cosmic expansion rates appearing in the Taylor expansion, when compared to the previous analyses. We also note that successive terms in the series  decrease sufficiently fast, thereby verifying  the assumption of  Taylor expansion.  It is expected that in the near future, as the SN dataset becomes large enough, these coefficients  get sharply peaked marginal likelihoods and  become the most basic model-independent description of the expansion history of the universe.

\section{Bayesian model-independent approach}

For parameterized models with parameters $\alpha , \beta , ..$,  the likelihood for the model $M_i$ or the Bayesian evidence [denoted as ${\cal L}(M_i)$], which is the probability for the data $D$ given the truth of the model $M_i$,
can be evaluated as

\begin{eqnarray}\label{eq:likelimod}
& p(D|M_i,I)    \equiv {\cal L}(M_i) \nonumber     \\ 
            &  =\int d\alpha \int d\beta ... p(\alpha , \beta ,
...|M_i) {\cal L}_i (\alpha , \beta , ...), 
\end{eqnarray}
where $p(\alpha , \beta , ...|M_i)$ is the prior probability for the set of
parameter values $\alpha , \beta , ..$ given the truth of $M_i$ and ${\cal L}_i (\alpha , \beta , ...)$ 
is their likelihood function.
 The likelihood function is often taken  to
be \cite{drell}

\begin{equation}\label{eq:likelipar}
{\cal L}_i (\alpha , \beta , ...)=\exp \left[-\chi _i^2(\alpha , \beta , ..)/2
\right].
\end{equation}
where $\chi_i^2$ is the $\chi^2$-statistic. Another quantity of interest is the marginal likelihood for any one parameter, say $\alpha$,  in a model. This is obtained by integrating the  integrand in (\ref{eq:likelimod}) over all parameters, except $\alpha$. Thus in the above case, the marginal likelihood for the parameter $\alpha $ can be obtained as

\begin{equation}
{\cal L}_i(\alpha)=\int d\beta \int d\gamma ... p(\alpha , \beta ,\gamma
...|M_i) {\cal L}_i (\alpha , \beta , \gamma ...). \label{eq:marginal}
\end{equation}

While using the apparent magnitude-redshift data, the observable is the distance modulus $
\mu =5 \log \left({D}/ {1 \hbox {Mpc}}\right)+25
$, a function of redshift $z$ and contains parameters $\alpha$, $\beta$, etc. in the model concerned.  $D(z; \alpha , \beta ..)$ is called the luminosity distance and  $D/1$ Mpc indicates that it is expressed in units of megaparsec.

 In the model-independent analysis of SN data  \cite{mvj1,mvj2}, the scale factor of the universe is expanded into a  Taylor series in $t$ about the present epoch $t_0$. 
With $t-t_0\equiv T$, where $t_0$ is the present time,  the Taylor series can be written as

\begin{eqnarray}
 & a(t_0+T) = a_0 \times   \\ \nonumber
 & \left[1+H_0T-\frac{q_0H_0^{2}}{2!}T^2 
  +\frac{r_0H_0^3}
                    {3!}   T^3  
   -\frac{s_0H_0^4}{4!}T^4 +\frac{u_0H_0^5}{5!}T^5+.. \right] 
         \label{eq:taylor}
\end{eqnarray}
Note that $T$ assumes negative values. Here the parameters in the theory are  the present value of the scale factor $a_0$, the Hubble parameter $H_0\equiv 100 h$ km s$^{-1}$ Mpc${-1}$ , the deceleration parameter $q_0$, higher order expansion rates such as $r_0$, $s_0$, $u_0$, etc. and the curvature scalar $k=\pm 1$. (The spatially flat $k=0$ case is incorporated by including sufficiently large values of $a_0$.)  One of our tasks is to deduce the values of these parameters  from the observational data.

For  a light pulse   emitted  from an SN situated at the coordinate $r_1$ at time $t_1$ and reaching us at $r=0$ at time $t_0$,  the RW metric allows one to write

\begin{equation}
\int_{t_1}^{t_0} \frac{cdt}{a(t)} = \int_{r_1}^{0} \frac{dr}{(1-kr^2)^{1/2}} \label{eq:integral}.
\end{equation}
For a $k=0$ RW metric,  this  can be used
 to obtain

\begin{equation}
r_1= \int_{t_1}^{t_0} \frac{cdt}{a(t)}=\int_{T_1}^{0}\frac{dT}{a(t_0+T)}.
\end{equation}
Similar expressions can be found for $k=\pm 1$ cases too. With this, we may
 compute the luminosity distance $D=r_1a_0(1+z)$.
An important part of the calculation is the solution of the following equation, used to find $T_1$ in terms of $z$, for each combination of parameter values. This is done in a direct and purely numerical way:

\begin{equation}
 1+z=\frac{a(t_0)}{a(t_0+T_1)} \label{eq:numsolzT}
\end{equation} 
We may thus obtain the distance modulus $\mu=5 \log \left({D}/{1 \hbox {Mpc}}\right)+25$.
Here $D$ and hence $\mu$ are functions of  $z$ and  contain parameters  $k$, 
$a_0$,
$H_0$,
$q_0$,
$r_0$, $s_0$, $u_0$, etc.

The likelihood function is now $
{\cal L}= \exp[-\chi^2(k,h,a_0,q_0,r_0,s_0,u_0)/2]
$
where $\chi^2$ is given by

\begin{equation}\label{eq:chi2}
\chi^2 = \Sigma _k \left( \frac{\hat{\mu}_k -\mu_k(z_k;k,h,a_0,q_0,r_0..)}{\sigma_k}\right)^2.
 \end{equation}
 Here $\hat{\mu}_k$ is the measured value of the distance modulus  of the $k^{th}$ supernova, $\mu_k(z_k;k,h,a_0,q_0,..)$ is its expected value (from theory) and $\sigma _k$ is the uncertainty in the measurement.

The likelihood  for the truncated Taylor series form of scale factor can be found using equation (\ref{eq:likelimod}) as

\begin{eqnarray} \nonumber
& {\cal L}(M_i) =\frac{1}{2}\sum_{k=-1,1} \int dh \int da_0\ \int dq_0\ \int dr_0\ \int ds_0\ \int du_0 \\
&p(h) p(a_0)p(q_0)p(r_0)p(s_0)p(u_0)\; 
e^{-\chi^2 /2} . \label{eq:likelihood}
\end{eqnarray}
where $p(h)p(a_0)p(q_0)p(r_0)p(s_0)p(u_0)$  is a product of Gaussian  probability distributions evaluated using the mean values and standard deviation of the marginal likelihoods  obtained in  previous analyses [\cite{mvj2} in our case] for each of the parameters. This is an approximation to $p(h,a_0,q_0,r_0.....\mid M_i)$, the prior probability to be introduced  in equation (\ref{eq:likelimod}).

In the above, we have kept terms up to fifth order in the  Taylor expansion.  Increasing the number of terms by unity will enhance the computation time  by more than an order of magnitude. We have performed this computation  with various orders in the truncated series, starting with  second order in which only the first three terms are kept. For  orders different from 5, necessary changes are to be made in the above expressions. 

\section{Marginal likelihoods for the cosmic expansion rates}

We first obtain the marginal likelihoods  for the various expansion rates in the case of truncated Taylor series of order 5. This is  a repetition of the calculation in \cite{mvj1,mvj2}, using the Constitution data. The marginal likelihoods we obtain here  give valuable information regarding the expansion history of the universe.

An important step  made in the present computation of marginal likelihood is that while using (\ref{eq:likelihood}), the marginal likelihoods in the previous analysis \cite{mvj1,mvj2} are taken as the prior probability distributions, for the corresponding coefficients.  References \cite{mvj1,mvj2} have used  flat priors, since there were no other previous work evaluating these marginal likelihoods. But there itself it was proposed that the posterior marginal likelihoods obtained shall be used for subsequent analysis and the present work is the appropriate place to make use of this. However, it would not be computationally feasible  to  use  the posterior in the previous analysis  as prior in terms of a table of values; instead, as stated above, we approximate those  distributions by Gaussian functions with the corresponding mean and standard deviations obtained in \cite{mvj2}. A comparison with the actual plots show that this is a reasonable approximation for most coefficients and at any rate is a better option than flat priors. The product of such individual priors is the combined prior,  used in equation (\ref{eq:likelihood}).

 In the present model-independent analysis,  the `Constitution'  data \cite{hicken} of 397 SN were used. We have computed the marginal likelihoods of four important   expansion rates of the present universe, namely  $q_0$, $r_0$, $s_0$ and $u_0$, and the results are shown in Figs. (1)-(4). We have kept terms up to fifth order in this computation, but only the flat ($k=0$) case is considered here. This is equivalent to assuming a $\delta$-function prior for the flat spatial geometry. 
The joint prior probability used for other parameters was, as described above,   the product of individual Gaussian functions in each parameter with mean and standard deviations as  follows: 
 $h=0.68\pm 0.06$, $q_0=-0.90\pm 0.65$, $r_0=2.7\pm 6.7$, $s_0=36.5\pm 52.9 $, and $u_0=142.7\pm 320$ \cite{mvj2}. In each case, the integrations were performed in the 2$\sigma$ range of each of the parameters. We have performed variation with respect to $h$, though marginal likelihood for this parameter was not drawn. The step sizes chosen for these parameters were $\Delta  h=0.01 $, $\Delta q_0 =0.1 $, $\Delta r_0 =1 $, $\Delta s_0 =20 $ and $\Delta u_0 = 100$.

\FIGURE{
\includegraphics{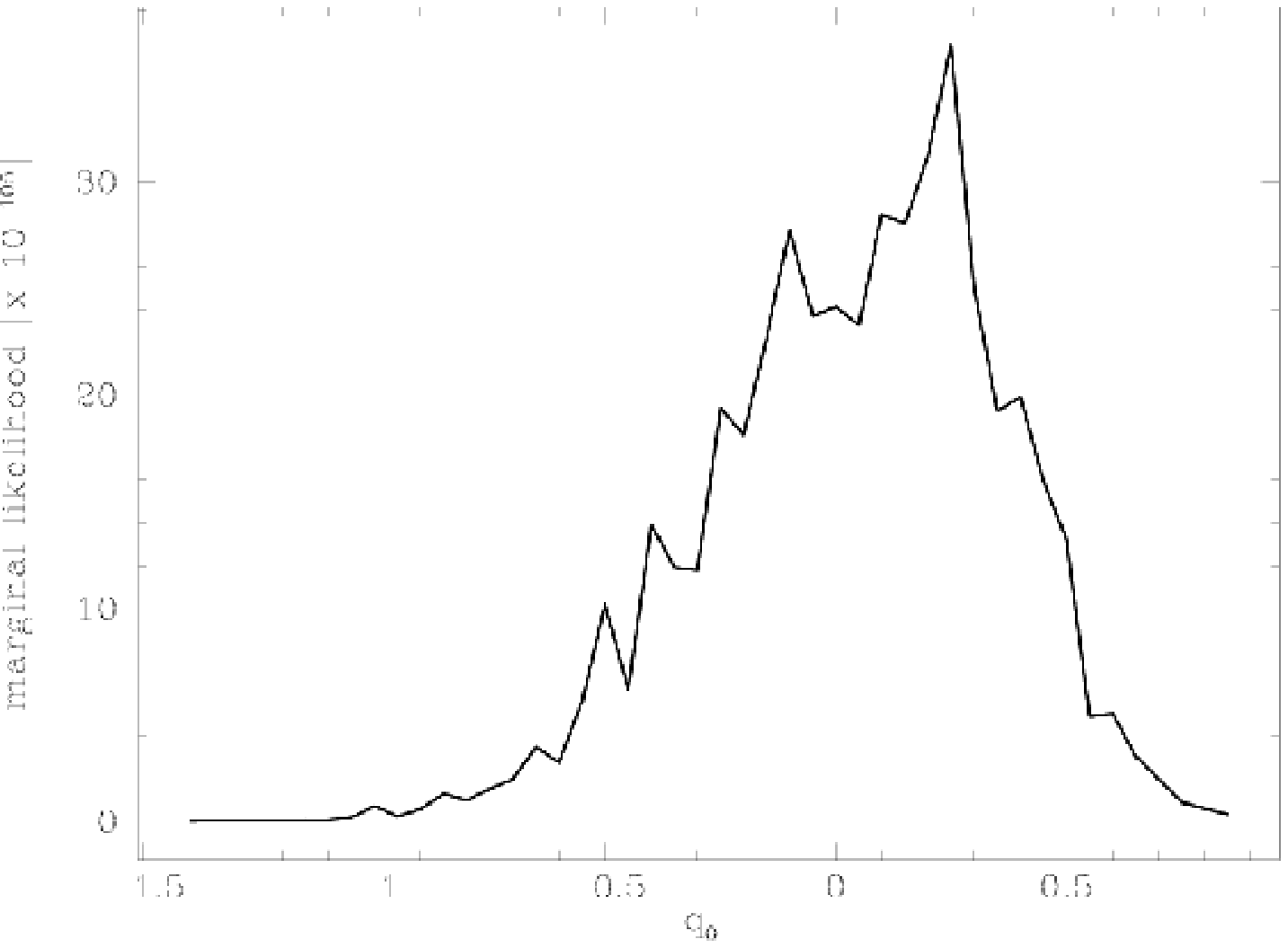}
\caption{\label{fig:fig1} Marginal likelihood for the parameter $q_0$ (in units of $10^{-105}$), while using the polynomial of order 5. }
}

\FIGURE{
\includegraphics{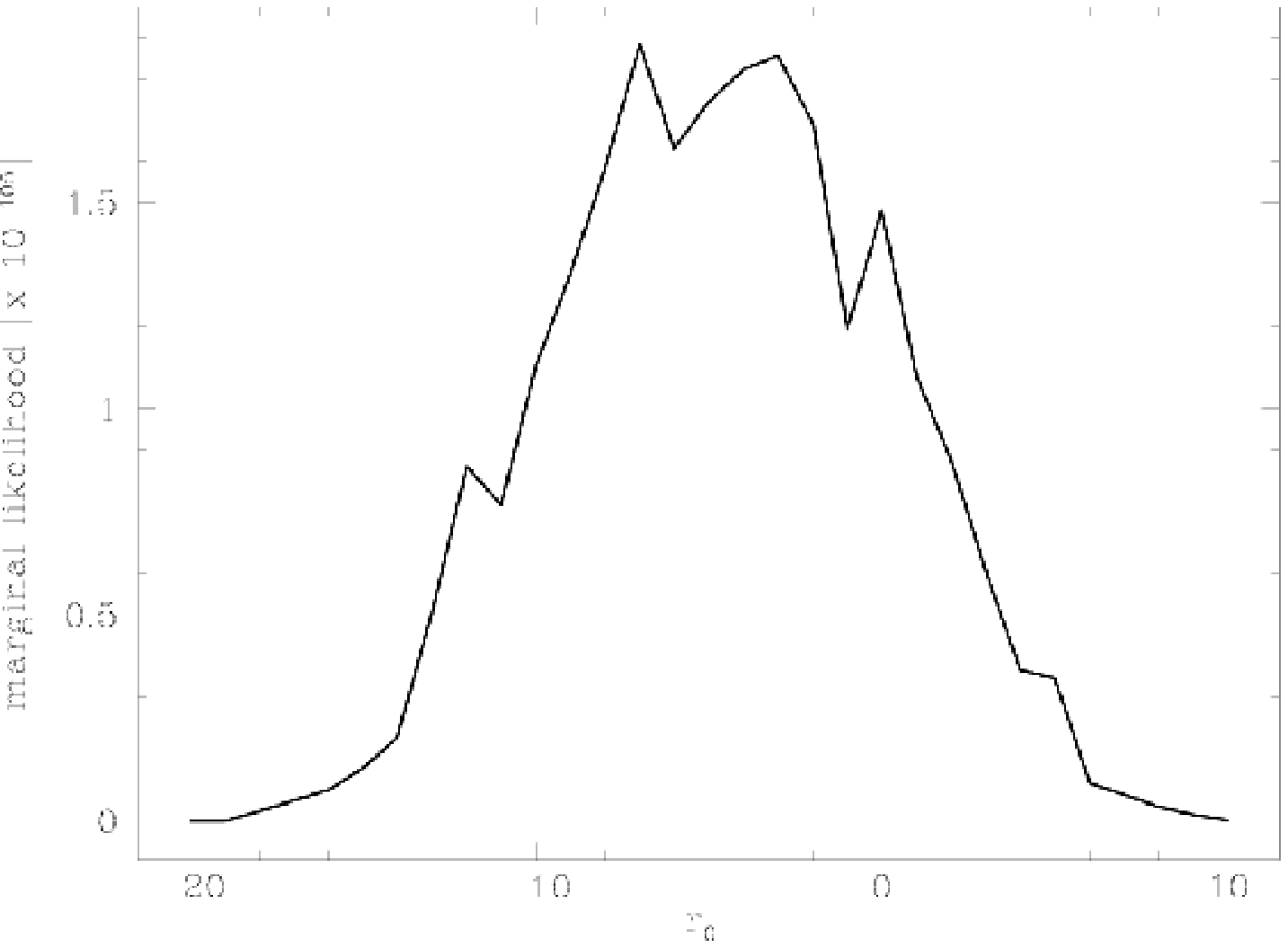}
\caption{\label{fig:fig2} Marginal likelihood for the parameter $r_0$ (in units of $10^{-105}$), while using the polynomial of order 5. }
}

\FIGURE{
\includegraphics{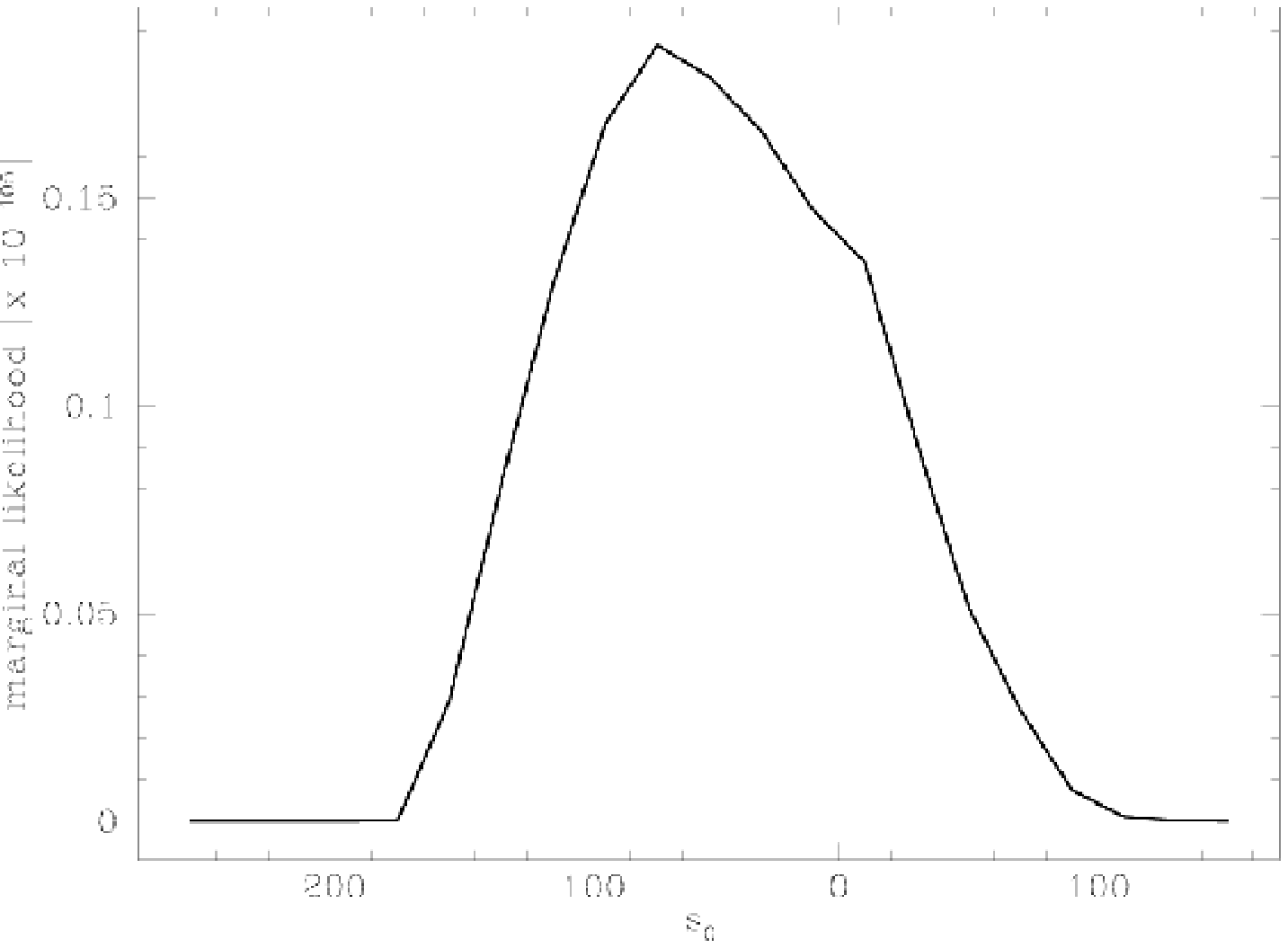}
\caption{\label{fig:fig3} Marginal likelihood for the parameter $s_0$ (in units of $10^{-105}$), while using the polynomial of order 5. }
}

\FIGURE{
\includegraphics{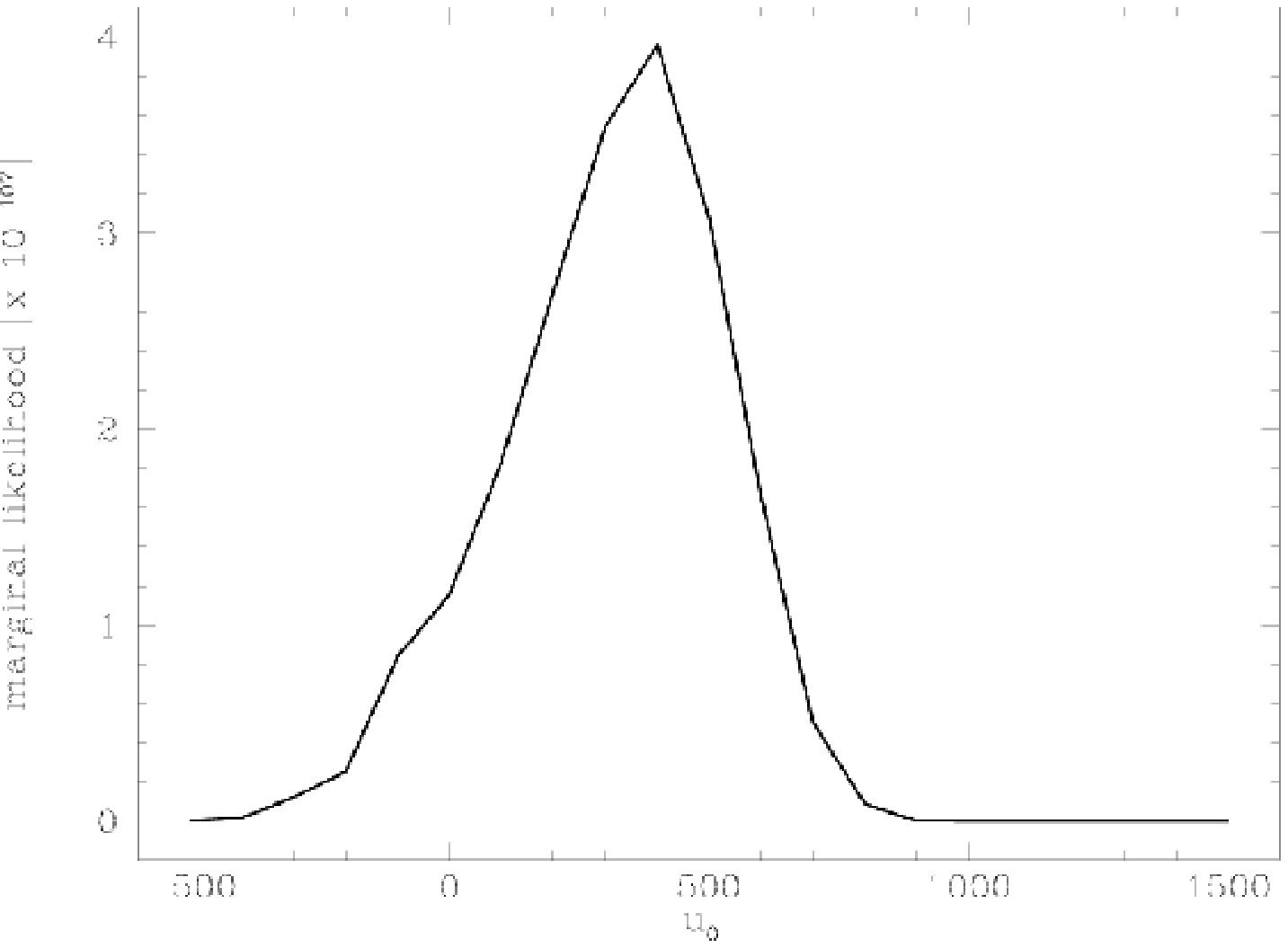}
\caption{\label{fig:fig4} Marginal likelihood for the parameter $u_0$ (in units of $10^{-107}$), while using the polynomial of order 5. }
}

\FIGURE{
\includegraphics{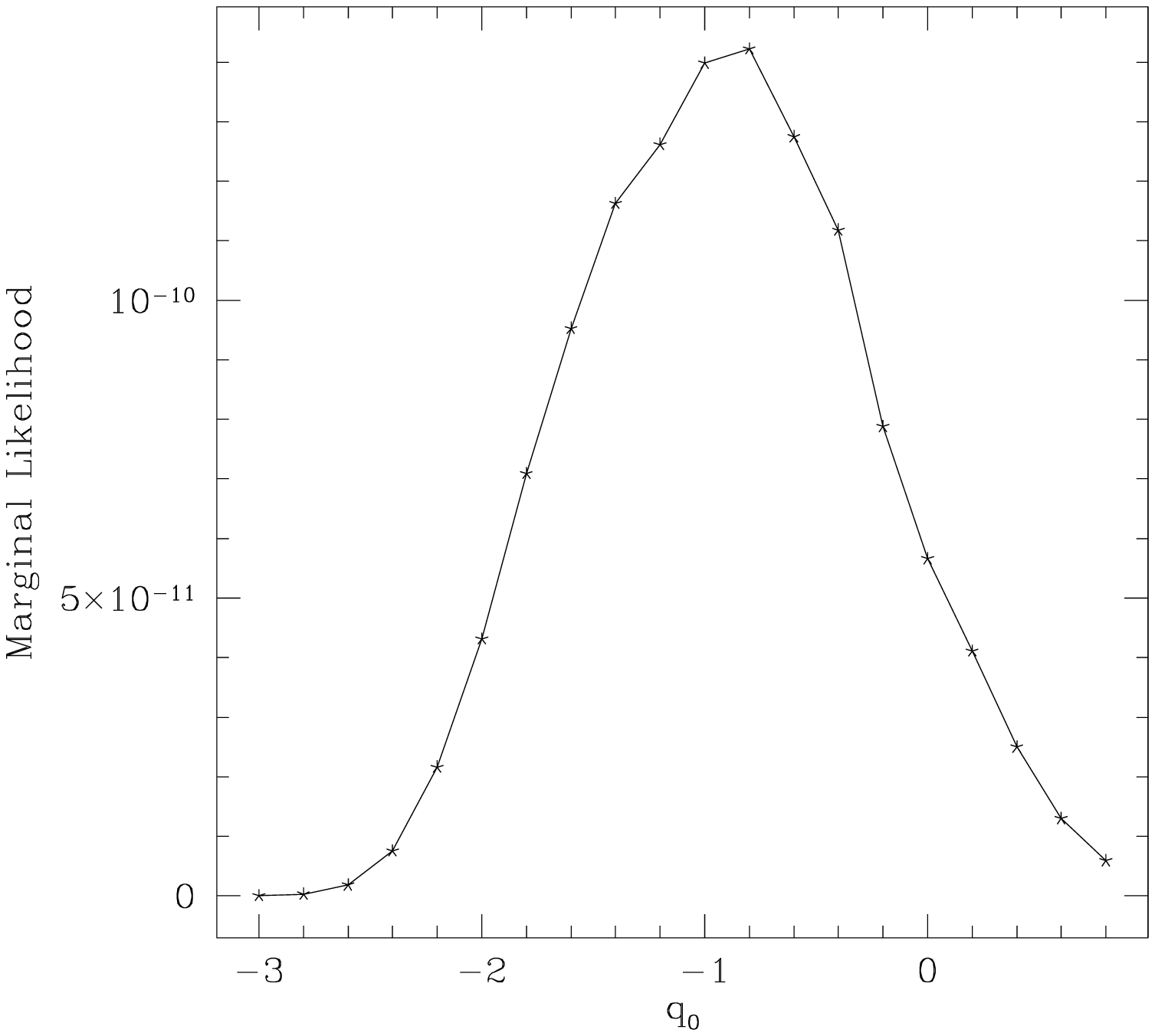}
\caption{\label{fig:fig5} Marginal likelihood for the parameter $q_0$ obtained in \cite{mvj2}, while using the polynomial of order 5 and the data in \cite{perl1}. }
}

The results  show that there is  significant constraining of the parameters  while using the new and refined data,  compared to the corresponding results in \cite{mvj2}. It is to be reminded that the marginal likelihoods are not precisely probability distributions for the parameters; instead, they are the probability for the data, given the model and the parameter values. However, we here compute mean and standard deviations considering them as distributions. The new mean and standard deviations are the following:  $q_0=0.04\pm 0.30$, $r_0=-4.5\pm 4.6$, $s_0=-42.8\pm 52.5 $, and $u_0=320.5\pm 213.0$. The marginal likelihood for $q_0$ obtained in \cite{mvj2} [which was not much different from that of \cite{mvj1}, both using the same data in \cite{perl1}],  is reproduced here in Fig. (\ref{fig:fig5}) for comparison with the distribution in Fig. (\ref{fig:fig1}). It can be seen that the standard deviations of each of these parameters except that of $s_0$ have decreased substantially, which leads to our above assertion.

It shall be noted that even when beginning with a  prior centred around $q_0=-0.9$, which is strongly in favor of an accelerated expansion, we ended up with this marginal likelihood peaked around $q_0\approx 0$ slightly towards the positive side. This casts doubt on the paradigm of cosmic acceleration itself. Where as the data in \cite{perl1}  validated the claim of accelerated expansion  \cite{mvj1,mvj2}, the new extended SN  dataset in \cite{hicken} indicates that the universe  is neither accelerating nor decelerating. This agrees with the analysis in \cite{varun}, which finds that a `coasting' ($q_0=0$) evolution for the universe is equally plausible. However, the presence of substantial amount of dark energy and dark matter would still be required to explain the data.

The considerable spread in the marginal likelihoods  shows that there is still enough freedom in choosing the values of those parameters for a best fit.  In other words, even now there is a sizable volume in the parameter space that can have the same low $\chi^2$. But  this should not be viewed as a drawback of the analysis; instead, this simply reflects the fact that the data are  not yet accurate enough.
Several recent analyses of Constitution SN data have reported that such freedom exists \cite{varun,wei}.   This freedom in SN data was noted  earlier in \cite{mvj1,mvj2}, which highlights the strength of the Bayesian model-independent approach.

Based on the mean values obtained for these parameters, we compute the successive terms in the series (\ref{eq:taylor}). With time in units of $10^{17}$ s, the series can be written as

\begin{eqnarray} \nonumber
 1+2.106\times 10^{-1} T-2.22\times 10^{-2} q_0 T^2+1.55\times 10^{-3}r_0 T^3 \\
 - 0.819\times 10^{-4} s_0 T^4 +3.45\times 10^{-6} u_0 T^5+......
  \label{eq:converge}
\end{eqnarray}
where we have taken $h=0.65$. With the values of the parameter in the ranges obtained in the analysis, this series appears to converge even for  $\mid T \mid$ as large as $\approx 3\times 10^{17}$ s. However, this feature is not  essential for our analysis, for we have assumed only a polynomial form for the scale factor. The situation was not different in \cite{mvj1,mvj2} either. 

\section{ Likelihoods for the models and the model-independent approach}

Now we evaluate the likelihoods for various models and also for the model-independent approach. The realistic models  considered are the flat LCDM model with constant equation of state parameter $w=-1$ and the more general flat dark energy model with the CPL equation of state. For the model-independent case, we have used truncated serieses of various orders, starting with order 2 to the largest order 6 for the scale factor. We have not restricted ourselves to the flat case in this analysis (whereas in the above section, all the marginal likelihoods were computed by assuming a fifth order polynomial and with $k=0$).

These  likelihoods in different models are important since the Bayes factor is the ratio between them. We consider that there is positive evidence for a model, when compared with another one, only if its likelihood  is greater than 3 times the value of the latter \cite{drell}. The evidence is considered to be strong only if its likelihood is larger than 20 times and very strong if it is larger than 150 times \cite{drell} that of the other. Therefore it is highly desirable to find the likelihoods or Bayesian evidences for  models and we now compute them for the cases mentioned above.

   The likelihood for polynomials of various orders we have found are given in Table 1.
It is  seen that in this  model-independent approach, the polynomial of order 4 has the largest likelihood $\approx 770 \times 10^{-105}$ and as discussed in the introduction, this value shall be taken as the likelihood for the model-independent approach.

\TABLE[pos]{
 \centering
 \begin{minipage}{85mm}
  \caption{Likelihoods/Bayesian evidence for various models. We have taken   $k=0, \pm 1$ in the case of polynomial approximations but $k=0$ for the  two realistic models,.}
  \begin{tabular}{ll} \hline
Polynomial of order 2 (up to $q_0$) &$350 \times 10^{-105}$ \\
Polynomial of order 3 (up to $r_0$)&$ 90\times 10^{-105}$\\
Polynomial of order 4 (up to $s_0$)&$ 770 \times 10^{-105}$\\
Polynomial of order 5 (up to $u_0$)&$30 \times 10^{-105}$\\
Polynomial of order 6 (up to $v_0$)&$10\times 10^{-105}$ \\ \hline
Flat LCDM model & $9.8\times 10^{-105}$ \\ \hline
Flat dark energy model & $0.89\times 10^{-105}$\\ 
 with CPL equation of state &  \\ \hline
\end{tabular}
\end{minipage}

}

For flat LCDM model, the likelihood was $9.8\times 10^{-105}$ and for flat dark energy model with CPL equation of state, it was $0.89\times 10^{-105}$. These  values, when compared to the  likelihood $770 \times 10^{-105}$ of the model-independent approach,  show that the performaces of those realistic models are not as good as the polynomials.

The priors used for the coefficients in the polynomial form of the scale factor are the same as the Gaussian priors used in the previous section. For $v_0$, we took a flat prior in the range $-3000<v_0<3000$ in the same manner as we chose flat priors for other parameters in the previous work; i.e., flat priors only for the contributing regions of the parameter concerned. For the realistic models, we took $k=0$ and the prior for $h$ was the same as that in the previous section. $\Omega_m$ had Gaussian prior, with mean and standard deviations  $\Omega = 0.29\pm 0.025$. Integrations were performed over $2\sigma$ range of these parameter values.

The  two additional parameters in the  model with CPL equation of state were chosen to have  flat priors with $-1.2<w_0 <0$, $-12<w_1<0$, which are  judged from some recent computations \cite{wei}. Variations in these priors can change the likelihoods for this model to some extent, which can  affect our conclusions regarding the dark energy model in a similar manner. But we  note that  the lowest value for $\chi^2$ claimed by \cite{wei} for this model is 461.254 at $\Omega_m = 0.453$, $w_0=-0.207$ and $w_1=-11.316$. Even if we set $\delta$-function priors  at these parameter values, the likelihood can only reach the maximum possible value  $=6.9\times 10^{-101}$, which  corresponds to this $\chi^2$ value. But  the Gaussian prior peaked at $\Omega_m=0.29 $  we must use is far from such a $\delta$-function distribution located at $\Omega_m = 0.453$, so that there is no chance  for this model  getting  a high value for the likelihood. Moreover, using the same $\delta$-function distributions as priors for these parameters, while analyzing some future data, may turn out to be extremely harmful for this model. Therefore we resort to using the   flat priors for $w_0$ and $w_1$, as mentioned above, but at the same time caution that the present conclusion regarding the model with CPL equation of state is liable to change since we do not have fiducial priors for these parameters. Only after several repetitions can we reach such priors for them. However, the low value of likelihood obtained for the LCDM model is more or less robust.

The step sizes chosen for the parameters were $\Delta  h=0.01 $, $\Delta q_0 =0.1 $, $\Delta r_0 =1 $, $\Delta s_0 =20 $ and $\Delta u_0 = 100$, as in the previous case. For other parameters, we took $\Delta v_0 = 1000$ and $\Delta a_0 = 1000$. The flat case is incorporated here by including large values of $a_0$, say up to $a_0=8000$ Mpc and considering only $k=\pm 1$. For $w_0$ and $w_1$, we chose $\Delta w_0 =0.1 $ and $\Delta w_1 =1 $.

\section{Conclusion}
We assumed that  a Taylor series form for the unknown variable $a(t)$, which describes the cosmic expansion history,  is valid and first attempted to find the coefficients in this expansion using the recent Constitution SN data. It is found that there is significant constraining of these parameters when compared to a previous analyses using the  data in \cite{perl1}. The new marginal likelihoods for various coefficients evaluated with the Constitution data  lead us to expect that when more refined and abundant SN data set becomes available in the near future, the curves will get sharply peaked and this method of direct determination of the cosmic expansion history shall prove to be indispensable.

One of the notable results obtained from the present analysis  is the shift in the computed mean value of the deceleration parameter $q_0$, from that found in the previous analysis. Even when we start with a  prior probability distribution that strongly favors an accelerating universe,  the present analysis using Constitution data provide a marginal likelihood peaked around the zero of the deceleration parameter.  This result is in  agreement with the analysis in \cite{varun}. However, we reiterate that the considerable spread still found in the likelihoods of these parameters indicate  freedom in the choice of their numerical values. This also is in agreement with several other analyses of the same data using alternative methods \cite{varun,wei}.  

Another  attempt we made in this paper is the comparison of the performance of some realistic cosmological models with that of the model-independent approach, in explaining the SN data. 
We note that a great practical use of scientific theories is that the equations they provide save us from  keeping large amount of raw data, for use in future  applications. Therefore it is natural to expect that the likelihood for   a fundamental scientific theory  exceed that of a truncated Taylor series of the unknown function (the model-independent approach, as we refer to in this paper) or at least be equal to that of such a series. We here find that  the two popular realistic cosmological models  analyzed are much behind  a Taylor expansion for the scale factor  in explaining the data, since their Bayesian evidences/likelihoods  are  less than that in the latter case. We have used fiducial Gaussian priors for $\Omega_m$ in the flat LCDM model, but caution that the likelihood for the dark energy model with CPL equation of state is liable to change since such priors are not available for the parameters $w_0$ and $w_1$.

A distinguishing feature of our analysis is that the marginal likelihoods for each parameter obtained in the previous case is chosen as the prior probability distribution in the present one, thereby implementing the Bayesian method in true spirits. The work is also intended as a demonstration of this fundamental requirement in Bayesian analysis.
We have also verified that successive terms in the series decreases fast enough, justifying the Taylor expansion hypothesis, though this is not essential since our basic assumption  was that the scale factor is expressible as a polynomial.

What we envisage here is a combined Bayesian model-independent approach. The application of this method answers one of the criticisms raised against Bayesian model comparison - that it is soft towards  models with poor explanatory power, since it only compares the available ones. We now have a new quantity, which we call likelihood for the model-independent approach. Evaluation of this likelihood/evidence is only a logical extension of the evaluation of marginal likelihoods of those coefficients in the expansion and it is now demonstrated that this quantity can be used to calibrate the performace of cosmological models.

\acknowledgments{
It is a pleasure to thank Professor J. V. Narlikar for helpful discussions and  Asis,  Sandeep, Joe and Vivek for useful computing tips. The author also wishes to thank IUCAA, where most of these computations were done, for hospitality during a visit under the associateship program.}


\begin{thebibliography}{99}

\bibitem{per99} S. Perlmutter  {\sl et al.}, {\sl Measurements of Omega and Lambda from 42 high-redshift supernovae}, 1999 {\sl Astrophys. J. }
  {\bf 517}, 565 [arXiv:astro-ph/9812133]
\bibitem{rie98} A. G. Riess {\sl et al.},  {\sl Observational evidence from supernovae for an accelerating universe and a cosmological constant}, 1998 {\sl   Astron. J.} {\bf 116},
  1009 [arXiv:astro-ph/9805201]
\bibitem{cpl1} M. Chevallier and D. Polarski,  {\sl Accelerating universes with scaling dark matter}, 2001 {\sl  Int. J. Mod. Phys. D} {\bf 10}, 213 [arXiv:gr-qc/0009008]
\bibitem{cpl2} E. V. Linder,  {\sl Exploring the expansion history of the universe}, 2003 {\sl  Phys. Rev. Lett.} {\bf 90}, 091301 [arXiv:astro-ph/0208512]
\bibitem{pavon} L. P. Chimento and D. Pavon,  {\sl Dual interacting cosmologies and late accelerated expansion}, 2006 {\sl  Phys. Rev. D} {\bf 73}, 063511 [arXiv:gr-qc/0505096] 
\bibitem {intrde}  D. Rowland and I. B. Wittingham,  {\sl Models of interacting dark energy}, {\sl  Mon. Not. R. Astron. Soc.} {\bf 390}, 1719 (2008).
\bibitem {gong} Y. Gong and  X. Chen,  {\sl Two component model of dark energy}, 2007 {\sl  Phys. Rev. D} {\bf 76}, 123007 [arXiv:0708.2977]
\bibitem{jaffe} A. Jaffe,  {\sl $H_0$ and odds on cosmology
}, {\sl  Astrophys. J.} {\bf 471}, 24 (1996).
\bibitem{drell}  P. S. Drell, T. J. Loredo and  I. Wasserman,  {\sl Type IA supernovae, evolution, and the cosmological constant}, 2000 {\sl  Astrophys. J.} {\bf 530}, 593  [arXiv:astro-ph/9905027] 
\bibitem{mvjvn}  M. V. John and   J. V. Narlikar,  {\sl Comparison of cosmological models using Bayesian theory}, 2002 {\sl 
 Phys. Rev. D}  {\bf 65}, 
043506 [arXiv:astro-ph/0111122] 
\bibitem {hobson} M. P. Hobson, S. L. Bridle and  O. Lahav,  {\sl Combining cosmological datasets: hyperparameters and Bayesian evidence}, 2002 {\sl  Mon. Not. R. Astron. Soc.} {\bf 335}, 377 [arXiv:astro-ph/0203259]
\bibitem{pia} P. Mukherjee  {\sl et al.},  {\sl Model selection as a science driver for dark energy surveys}, 2006 {\sl  Mon. Not. R. Astron. Soc.} {\bf 369}, 1725 [arXiv:astro-ph/0512484]
\bibitem{trotta}  R. Trotta,  {\sl Bayes in the sky: Bayesian inference and model selection in cosmology}, 2008 {\sl  Contemp. Phys.} {\bf 49}, 71 [arXiv:0803.4089] 
\bibitem{mvj1}  M. V. John,  {\sl Cosmographic evaluation of the deceleration parameter using type Ia supernova data}, 2004 {\sl  Astrophys. J.} {\bf 614}, 1 [arXiv:astro-ph/0406444]
\bibitem{mvj2}  M. V. John,  {\sl Cosmography, decelerating past, and cosmological models: Learning the Bayesian way}, 2005 {\sl  Astrophys. J.} {\bf 630}, 667 [arXiv:astro-ph/0506284]
\bibitem{turner} C. Shapiro and  M. S. Turner,  {\sl What do we really know about cosmic acceleration?}, 2006 {\sl  Astrophys. J.}  {\bf 649}, 563 [arXiv:astro-ph/0512586]
\bibitem{visser} C. Cattoen and M. Visser, {\sl Cosmography: Extracting the Hubble series from the supernova data}, 2007 [arXiv:gr-qc/0703122v3]
\bibitem{lima} A. C. C. Guimaraes, J. V. Cunha and J. A. S. Lima, {\sl Bayesian analysis and constraints on kinematic models from Union SNIa data}, 2009 {\sl JCAP} {\bf 10}, 010 [arXiv:0904.3550v3] [astro-ph.CO]
\bibitem{seikel} M. Seikel and  D. J. Schwarz,  {\sl Model- and calibration-independent test of cosmic acceleration}, 2009 {\sl  JCAP} {\bf 2}, 024 [arXiv:0810.4484]
\bibitem{varun}  A. Shafieloo,  V. Sahni and  A. A. Starobinski,  {\sl Is cosmic acceleration slowing down?},   2009 [arXiv:0903.5141v4][astro-ph.CO] 
\bibitem{wei}  H. Wei,  {\sl Tension in the recent type Ia supernovae datasets},  2009 [arXiv:0906.0828v1] [astro-ph:CO] 
\bibitem{hicken}  M. Hicken {\sl et. al.},  {\sl Improved dark energy constraints from ~100 new CfA supernova type Ia light curves}, 2009 {\sl  Astrophys. J.} {\bf 700}, 1097 [arXiv:0901.4804]
\bibitem{perl1}  R. A. Knop {\sl et al.},  {\sl New Constraints on $\Omega_M$, $\Omega_\Lambda$, and w from an independent set of eleven high-redshift supernovae observed with HST}, 2003 {\sl  Astrophys. J.} {\bf 598}, 102 [arXiv:astro-ph/0309368] 
\end{thebibliography}
\end{document}